\documentclass[12pt]{iopart}
\usepackage{iopams}
\begin{document}

\letter{Ionic fluids: charge and density correlations near  gas-liquid criticality}

\author{Oksana Patsahan\footnote[1]{To
whom correspondence should be addressed (oksana@icmp.lviv.ua)},
Ihor Mryglod\dag\ and Jean-Michel Caillol\ddag}

\address{\dag\ Institute for Condensed Matter Physics
of the National Academy of Sciences of Ukraine
1 Svientsitskii Str. 79011 Lviv, Ukraine}

\address{\ddag\ Laboratoire de Physique Th\'eorique
CNRS UMR 8627, B\^at. 210
Universit\'e de Paris-Sud 91405 Orsay Cedex, France}

\begin{abstract}
The correlation functions of an ionic fluid with  charge and size asymmetry are
studied within the framework of the random phase approximation. The results obtained for the charge-charge correlation function demonstrate that the second-moment Stillinger-Lovett (SL) rule is satisfied away from the gas-liquid critical point (CP) but not, in  general, at the CP. However in the special case of a model without size assymetry the  SL rules are satisfied even at the CP.
The expressions for the density-density and charge-density correlation functions valid far and close to the CP are obtained explicitely.
\end{abstract}


\submitto{\JPCM}

\maketitle

In recent years much attention has been focused on an issue of the criticality and phase transitions in ionic fluids. For reviews of the experimental and theoretical situation, see Ref.  \cite{weingartner}- \cite{stell1}. One of the most widely used theoretical models of Coulomb systems is the so-called "restricted primitive model" (RPM),  in which the ionic fluid is modelled as an electroneutral binary mixture of charged hard spheres of equal diameter immersed in a structureless dielectric continuum. Early studies \cite{stellwularsen} established that  the model has a gas-liquid critical point (CP), the location of which was determined by recent  simulations \cite{ydp}-\cite{caillol}. Due to the controversial experimental findings, the critical behavior of the RPM has also been under active debates  and  strong evidence  for an Ising universal class has been found by computer simulation \cite{caillol,luijten}  as well as by a recent theoretical study \cite{patsahan_mryglod}.
More recently, the effects of size- and/or charge-asymmetry on the phase diagram have  been studied theoretically \cite{netz_orland}-\cite{aqua_3} and by simulation \cite{Yan-Pablo,panagiotopoulos2}.

The  more relevant issue concerns the behaviour of the two-body charge-charge correlation function  $G_{DD}(r)$ near the CP where the density fluctuations diverge strongly.
Recently Aqua and Fisher \cite{aqua_1} using a class of exactly soluble spherical models studied both symmetric and nonsymmetric version of $1:1$ ionic models. They showed that in the former case the two-point charge correlations remain of short range and obey the Stillinger-Lovett (SL) rule near and at the critical point. Otherwise they found the divergence of the charge correlation length to be precisely the same as the divergence of the density correlation length. They also found that the SL rule  fails at criticality.

In this letter  we address an issue of the behavior of the  correlation functions of an ionic fluid far and near the CP. For this purpose we consider a charge- and size-asymmetric continuous model of  an ionic fluid.
According to  \cite{aqua_1}  the Fourier transform of the charge-charge correlation function of the model $\widetilde{G}_{QQ}(k)$  can be presented  as
\begin{equation}
\widetilde{G}_{QQ}(k)\simeq 0+\xi_{Z,1}^{2}k^{2}-\sum_{p\geq 2}(-1)^{p}\xi_{Z,p}^{2p}k^{2p},
\label{SL1}
\end{equation}
where  the first vanishing term results from electroneutrality (zeroth-moment SL condition) while the second term yields  the second-moment
SL rule with  $\xi_{Z,1}=\xi_{D}$  ($\xi_{D}$ is the Debye length).
We intend to answer the question  whether the SL rules are satisfied in the cases of both   non-symmetrical and symmetrical ionic fluids.

In order to fix our notations we first consider the case of a fluid made of two species $\alpha=1,2$. In the grand-canonical (GC) ensemble an equilibrium state is characterized by the inverse temperature $\beta=1/k_{B}T$ ($k_{B}$ and $T$ are Boltzmann constant and  temperature, respectively) and two dimensionless local chemical potentials $\nu_{\alpha}(\bf{r})=\beta ( \mu_{\alpha}-  \psi_{\alpha}(\bf{r}))$, where $\mu_{\alpha}$ is the ordinary chemical potential and $\psi_{\alpha}(\bf{r})$ is some external field acting on particles. Recall that the GC partition function $\Xi\left[\nu_{1},\nu_{2} \right] $ is a log-convex functional of the $\nu_{\alpha}$.
The truncated two-body correlation functions are defined as
\begin{equation}
\label{G}
G_{\alpha\beta}\left[\nu_{1},\nu_{2} \right] (1,2)=
\frac{\delta^{2} \ln \Xi\left[\nu_{1},\nu_{2} \right] }{\delta \nu_{\alpha}(1)
\delta \nu_{\beta}(2)} \; .
\end{equation}
The Legendre transform of $\ln \Xi$ defined as
\begin{equation}
\label{A}
\beta \mathcal{A}\left[ \rho_{1}, \rho_{2}\right] =\sup_{\nu_{1},\nu_{2}}
\left\lbrace \sum_{\alpha}\left\langle \rho_{\alpha}\vert \nu_{\alpha}\right\rangle
-  \ln \Xi\left[\nu_{1},\nu_{2} \right]\right\rbrace \; ,
\end{equation}
gives us the free energy. In eq.\ (\ref{A}) Dirac notations have been used, i.e.
\begin{equation}
\left\langle \rho_{\alpha}\vert \nu_{\alpha}\right\rangle \equiv
\int_{\Omega} d^{3}\mathbf{r} \; \rho_{\alpha}(\mathbf{r}) \nu_{\alpha}(\mathbf{r}) \; ,
\end{equation}
where $\Omega$ denotes the volume of the system. $\beta \mathcal{A}\left[ \rho_{1}, \rho_{2}\right]$ is a convex functional of the local densities $\rho_{\alpha}(1)$ and it is the Legendre transform of $\ln \Xi\left[\nu_{1},\nu_{2} \right]$, i.e.
\begin{equation}
\label{B}
\ln \Xi\left[\nu_{1},\nu_{2} \right] =\sup_{\rho_{1},\rho_{2}}
\left\lbrace \sum_{\alpha}\left\langle \rho_{\alpha}\vert \nu_{\alpha}\right\rangle
-  \beta \mathcal{A}\left[ \rho_{1}, \rho_{2}\right]\right\rbrace \; .
\end{equation}
$\beta \mathcal{A}\left[ \rho_{1}, \rho_{2}\right]$ is the generator of direct correlation
  functions. In particular the two-body direct correlation functions are defined as
 \begin{equation}
\label{C}
C_{\alpha\beta}\left[\rho_{1},\rho_{2} \right] (1,2)= -
\frac{\delta^{2} \beta \mathcal{A}\left[ \rho_{1}, \rho_{2}\right] }{\delta \rho_{\alpha}(1)
\delta \rho_{\beta}(2)} \; .
\end{equation}
The functions $G$ and $C$ are related by the Ornstein-Zernike (OZ) equation
\begin{equation}
C_{\alpha\beta}(1,2) G_{\beta\gamma}(2,3) = -\delta_{\alpha\gamma}\delta(1,3) \; ,
\end{equation}
where summation and integration over the repeated mute indices are meant.

We now consider the case of ionic mixtures, particles of species $\alpha=1$ are supposed to bear a charge $+q$ and those of species $\alpha=2$ bear an opposite charge $-Z q$ ($Z>0$). It is obviously convenient to introduce the total number and charge densities, $\rho_{N}$ and
$\rho_{Q}$, by the relations
\begin{equation}
\left(\begin{array}{c}
\rho_{N}\\
\rho_{Q}
\end{array}\right)={\mathcal{M}}
\left(\begin{array}{c}
\rho_{1}\\
\rho_{2}
\end{array}\right),
\qquad
\left(\begin{array}{c}
\rho_{1}\\
\rho_{2}
\end{array}\right)={\mathcal{M}}^{-1}
\left(\begin{array}{c}
\rho_{N}\\
\rho_{Q}
\end{array}\right)
\end{equation}
with
\begin{equation}
{\mathcal{M}}=
\left( \begin{array}{cc}
1 &1 \\
1 & -Z
\end{array} \right), \qquad
{\mathcal{M}}^{-1}=\frac{1}{1+Z}
\left( \begin{array}{cc}
Z &1 \\
1 & -1
\end{array} \right)
\end{equation}
For a homogeneous fluid we certainly have the charge neutrality condition $\rho_{Q}=\left\langle \widehat{\rho}_{Q}\right\rangle =0$ where the hat denotes the microscopic charge density and the brackets $\left\langle \ldots\right\rangle $ denote a GC average. The (truncated) correlations of  $\rho_{N}$ and
$\rho_{Q}$ will be defined as $G_{AB}(1,2)=\left\langle \widehat{\rho}_{A}(1)\widehat{\rho}_{B}(2)\right\rangle -
\left\langle \widehat{\rho}_{A}(1) \right\rangle  \left\langle \widehat{\rho}_{B}(2)\right\rangle $, where $A (B) =N,Q$.

It follows from equations given above that we can introduce two symmetric matrices of correlation functions, namely
\begin{equation}
G=
\left( \begin{array}{cc}
G_{11} &G_{12} \\
G_{21} & G_{22}
\end{array} \right), \qquad
\overline{G}=
\left( \begin{array}{cc}
\overline{G}_{NN} &\overline{G}_{NQ} \\
\overline{G}_{QN} & \overline{G}_{QQ}
\end{array} \right)
\end{equation}
which are related by the matricial equations
\begin{eqnarray}
G & = & \mathcal{M}^{-1} \; \overline{G} \;  \mathcal{M}^{-1} \\
\overline{G}&=& \mathcal{M} \;  G \;  \mathcal{M} \; .
\end{eqnarray}
Similarly we introduce the $C$ and the $\overline{C}$ as
\begin{eqnarray}
C & = & \mathcal{M} \; \overline{C} \;  \mathcal{M} \\
\overline{C}&=& \mathcal{M}^{-1} \;  C \;  \mathcal{M}^{-1} \; ,
\end{eqnarray}
in such a way that the form of OZ equation is preserved, i.e.
\begin{equation}
\label{OZ}
\overline{C}(1,3) \overline{G}(3,2) = -U \; \delta(1,2) \; ,
\end{equation}
where $U$ is the unit matrix.

We are interested in the behavior of $G_{AB}(r)$ in the critical region
at large $r$ or equivalently at small $k$ in Fourier space. This study will be made in the framework of the random phase approximation (RPA) defined by the closure relation
\begin{equation}
\label{RPA}
C=C_{\rm{HS}} -\beta \Phi -\beta \Phi_{c} \; .
\end{equation}
In eq.\ (\ref{RPA}) $C_{\rm{HS}}$ is the matrix of the direct correlation functions of some reference system chosen here to be a hard sphere fluid (HS) mixture (with diameters $\sigma_{\alpha}$, $\alpha=1,2$). The matrix $\Phi$ is built from short range pair interactions and, finally, $\Phi_{c}$ denotes the matrix of Coulomb interactions. With these notations, the OZ equation\ (\ref{OZ}) can be rewritten in Fourier space as
\begin{eqnarray}
&&\left( \begin{array}{cc}
\widetilde{G}_{NN}(k) &\widetilde{G}_{NQ}(k) \\
\widetilde{G}_{QN}(k) & \widetilde{G}_{QQ}(k)
\end{array}\right) \\ \nonumber
&&=
-\left( \begin{array}{cc}
\widetilde{C}_{NN}^{\rm{HS}}(k)- \beta\widetilde{\Phi}_{NN}(k)\;&\widetilde{C}_{NQ}^{\rm{HS}}(k)- \beta\widetilde{\Phi}_{NQ}(k) \\
\widetilde{C}_{NQ}^{\rm{HS}}(k)- \beta\widetilde{\Phi}_{NQ}(k)\; & \widetilde{C}_{QQ}^{\rm{HS}}(k)- \beta\widetilde{\Phi}_{QQ}(k)+\displaystyle\frac{4\pi \beta q^{2}}{k^{2}}
\end{array} \right)^{-1},
\end{eqnarray}
where
\begin{eqnarray}
\widetilde{\Phi}_{NN}(k)&=&\frac{1}{(1+Z)^{2}}\left[
Z^{2}\widetilde{\Phi}_{11}(k) +
2 Z \widetilde{\Phi}_{12}(k)
+\widetilde{\Phi}_{22}(k)                            \right] \\
\widetilde{\Phi}_{NQ}(k)&=&\frac{1}{(1+Z)^{2}}\left[
\widetilde{\Phi}_{11}(k) -2\widetilde{\Phi}_{12}(k) +  \widetilde{\Phi}_{22}(k)\right] \\
\widetilde{\Phi}_{QQ}(k)&=&\frac{1}{(1+Z)^{2}}\left[
Z \widetilde{\Phi}_{11}(k) +(1-Z)\widetilde{\Phi}_{12}(k)
-  \widetilde{\Phi}_{22}(k)\right] \,
\end{eqnarray}
and the tildes mean a Fourier transform.

A general study of the limit $k \to 0$ of eq.\ (\ref{RPA}) would be tedious and we make the following  simplifying assumptions:
\begin{itemize}
\item The HS will all be of the same diameter $\sigma_{\alpha}=\sigma$.
\item We split the potential $\widetilde{\Phi}_{\alpha\beta}(k)$ into two parts, i.e. $\widetilde{\Phi}_{\alpha\beta}(k) = \widetilde{\Phi}_{\alpha\beta}^{R}(k) +\widetilde{\Phi}^{A}(k)$ where $\widetilde{\Phi}_{\alpha\beta}^{R}(k)>0$ denotes the repulsive part and $\widetilde{\Phi}^{A}(k)<0$  denotes the attractive one. The attractive part   $\Phi^{A}$ is necessary to induce a liquid-vapor transition in the RPA approximation. Since we have considered HS of equal diameters the repulsive potentials $\widetilde{\Phi}_{\alpha\beta}^{R}(k)$ are used to mimic the soft core asymmetric repulsive interactions.
\end{itemize}
Moreover, at small $k$, we  assume that
\begin{eqnarray}
\widetilde{\Phi}_{\alpha\beta}^{R}(k) &= &\widetilde{\Phi}_{\alpha\beta}^{R}(0)
\left( 1 - \left(  b_{\alpha\beta}^{R} \; k\right)^{2} \right) + \mathcal{O}(k^{4}) \\
\widetilde{\Phi}^{A}(k) &= &\widetilde{\Phi}^{A}(0)
\left( 1 - \left(  b^{A} \; k\right)^{2} \right) + \mathcal{O}(k^{4}) \; ,
\end{eqnarray}
which implies that
\begin{equation}
\widetilde{\Phi}_{AB}(k) = \widetilde{\Phi}_{AB}(0)
\left( 1 - \left(  b_{AB} \; k\right)^{2} \right) + \mathcal{O}(k^{4}), \\
\end{equation}
where  $b_{AB}$ are complicated functions of $Z$, $b_{\alpha\beta}^{R}$, $b^{A}$, $\widetilde{\Phi}_{\alpha\beta}^{R}(0)$ and $\widetilde{\Phi}^{A}(0)$ which will not be displayed here. The $b_{\alpha\beta}^{R}$ can be interpreted as the effective diameters for the ions.

As a consequence of these assumptions we have a simple result
\begin{eqnarray}
\label{toto}
\left( \begin{array}{cc}
\widetilde{G}_{NN}(k) &\widetilde{G}_{NQ}(k) \\
\widetilde{G}_{NQ}(k) & \widetilde{G}_{QQ}(k)
\end{array}\right)
\\ \nonumber
=
\left( \begin{array}{cc}
\displaystyle\frac{1}{\widetilde{G}_{\rm{HS}}(k)}+ \beta\widetilde{\Phi}_{NN}(k)\;&\beta\widetilde{\Phi}_{NQ}(k) \\
\beta\widetilde{\Phi}_{NQ}(k)\; & \displaystyle\frac{1}{\rho Z}+ \beta\widetilde{\Phi}_{QQ}(k)+\displaystyle\frac{4\pi \beta q^{2}}{k^{2}}
\end{array} \right)^{-1},
\end{eqnarray}
where $\widetilde{G}_{\rm{HS}}(k)$ is the Fourier transform of the truncated two-body correlation function of a one-component HS fluid at the density $\rho_{N}$. We are now in position to study eq.\ (\ref{toto})  in the limit $k\to 0$.

At $k=0$ one finds that
\begin{equation}
\widetilde{G}_{NN}(0)=\frac{\widetilde{G}_{\rm{HS}}(0)}{1+\beta\widetilde{\Phi}_{NN}(0)\widetilde{G}_{\rm{HS}}(0)}
\end{equation}
or
\begin{equation}
\widetilde{G}_{NN}(0)=\frac{1}{\nu_{1}(\rho)+\beta\widetilde{\Phi}_{NN}(0)},
\end{equation}
where $\nu_{1}(\rho)=\partial \nu_{\rm{HS}}/\partial \rho$.

It is worth noting that if $\widetilde{\Phi}_{NN}(0) >0 $, then   $\widetilde{G}_{NN}(0)$ is  regular and there is no critical point (CP) in this case.  If $\widetilde{\Phi}_{NN}(0) <0 $, the isotherm compressibility $\chi_{T}=\beta \widetilde{G}_{NN}(0)/\rho^{2}$ can diverge signaling
the occurrence of a CP.  It turns out \cite{Cai-Mol} that $\nu_{1}(\rho)$ is a positive convex function of the density with a single minimum at $\rho_{c}\sigma^{3}=0.248$ with a value
$\nu_{1,\; c}=\nu_{1}(\rho_{c})=11.115$. It follows from this remark that, in the RPA approximation, the critical density is $\rho_{c}$ and that the critical temperature is given by $\beta_{c}\widetilde{\Phi}_{NN}(0)=
-\nu_{1, \; c}$. Hence, along the critical isochore and above $T_{c}$,   $\widetilde{G}_{NN}(0)$  behaves as
\begin{equation}
\widetilde{G}_{NN}(0)=\frac{1}{(\beta -\beta_{c})\widetilde{\Phi}_{NN}(0)}
\end{equation}
yielding the critical exponent of compressibility  $\gamma=1$ as expected.

Following  Aqua and Fisher \cite{aqua_1} we can cast the expressions for $\widetilde{G}_{NN}(k)$ and $\widetilde{G}_{QQ}(k)$ as
\begin{eqnarray}
\widetilde{G}_{NN}(k)&=& \frac{\widetilde{B}(k)}{\lambda_{2}(k)} +
\frac{1- \widetilde{B}(k)}{\lambda_{1}(k)} \\
\widetilde{G}_{QQ}(k)&=& \frac{\widetilde{B}(k)}{\lambda_{1}(k)} +
\frac{1- \widetilde{B}(k)}{\lambda_{2}(k)}
\end{eqnarray}
where $\lambda_{1}(k)$ and $\lambda_{2}(k)$ are the eigenvalues of the matrix of the direct correlation functions $\widetilde{C}_{AB}$ and
\begin{equation}
\widetilde{B}(k)=1 - \frac{{\widetilde\Phi}_{NQ}(0) k^{4}}{16 \pi^{2} q^{4}}+\mathcal{O}(k^{6}) \; .
\end{equation}
It is easily shown that at small $k$
\begin{equation}
\lambda_{2}(k)= (\beta -\beta_{c})\widetilde\Phi_{NN}(0) + a k^{2} +\mathcal{O}(k^{4}) \; ,
\end{equation}
with
\begin{equation}
a= -\frac{1}{2} \left. \frac{\partial ^{2} c_{\rm{HS}}(\rho_{c},k}{\partial k^{2}}\right \vert_{k=0} -\beta_{c}\widetilde\Phi_{NN}(0) b_{NN}^{2} -
\frac{\beta_{c}(\widetilde\Phi_{NQ}(0))^{2}}{4 \pi q^{2}} \; .
\end{equation}
In the above eq., $c_{\rm{HS}}$ is the ordinary direct correlation function connected to $C_{\rm{HS}}$ by $\widetilde{C}_{\rm{HS}}(k)=\widetilde{c}_{\rm{HS}}(k)  - 1/\rho$. Therefore, along the critical isochore above $T_{c}$ we have
\begin{equation}
\widetilde{G}_{NN}(k)=\frac{1}{(\beta -\beta_{c})\widetilde\Phi_{NN}(0)}\frac{1}{1 +
\xi^{2}k^{2}} \; \rm{ for}\; k \to 0 \; ,
\end{equation}
where the squared density-density correlation length is given by
\begin{equation}
\xi^{2}=\frac{a}{(\beta -\beta_{c})\widetilde\Phi_{NN}(0)} \; .
\end{equation}
The positivity of $\xi^{2}$, i.e. the positivity of  $a$, implies some restrictions on the various parameters of the model.
As expected, the critical exponent of the correlation length is $\nu=1/2$ and the Fisher exponent $\eta=0$. In the case where $q=0$ and $\Phi^{R}=0$ one recovers for $a$ the result of ref. \cite{Cai-Mol}.

Finally the behavior of $\lambda_{1}(k)$ for $k \to 0$ is found to be
\begin{equation}
\lambda_{1}(k)= \frac{4 \pi \beta q^{2}}{k^{2}}
\left( 1 +  \frac{k^{2}}{\overline{\kappa}_{D}^{2}}\right)
+ \mathcal{O}(k^{4}) \; ,
\end{equation}
where we have defined an effective squared Debye number as
\begin{equation}
\overline{\kappa}_{D}^{2}= \frac{\kappa_{D}^{2}}{1 + \beta\widetilde\Phi_{QQ}(0)\rho Z } \; ,
\end{equation}
$\kappa_{D}^{2}=4 \pi \rho \beta q^{2}Z$ being the usual squared Debye number.
Hence, once again along the critical isochore we get for $\widetilde{G}_{QQ}(k)$ the expression,
\begin{equation}
\widetilde{G}_{QQ}(k)=\frac{k^{2}}{4 \pi \beta q^{2}}
\left( 1 - \frac{k^{2}}{\overline{\kappa}_{D}^{2}}
\right)
+
\frac{({\widetilde\Phi}_{NQ}(0))^{2}k^{4}}{16 \pi^{2} q^{4}}
 \frac{1}{a(k ^{2} + \xi^{-2})}
+ \mathcal{O}(k^{6}) \;,
\end{equation}
which is valid far and at the CP at small $k$.

Some comments are in order. Firstly, away from the CP ($ \beta \neq \beta_{c}$) we have
$\widetilde{G}_{QQ}(k)\sim k^{2}/4 \pi \beta q^{2}$ which means that the both SL rules are satisfied.
This is in agreement with the results  obtained in \cite{evans}  for the case when the direct correlation function is given by (\ref{RPA}). In the general case, at the CP ($\beta= \beta_{c}, \rho= \rho_{c}$), $\xi^{-1}=0$ and we have for  $\widetilde{G}_{QQ}(k)$
\begin{equation}
\widetilde{G}_{QQ}(k)=k^{2}\left( \frac{1}{4 \pi \beta q^{2}}+\frac{(\widetilde\Phi_{NQ}(0))^{2}}{16 \pi^{2} q^{4}}\frac{1}{a}\right)+\mathcal{O}(k^{4}).
\end{equation}
As is seen from the above equation $\widetilde{G}_{QQ}(k=0)=0$ and thus the first SL rule is satisfied at the CP. By contrast, $\widetilde{G}_{QQ}(k\neq 0)$ is proportional to $k^{2}$ but with the coefficient that is not correct to ensure screening. Hence, the  second SL rule is violated at the CP.

However, in the special case of a symmetric model such that  $\Phi^{R}_{\alpha\alpha}=\Phi^{R}_{\alpha\beta}$ and thus $\widetilde{\Phi}_{NQ}(k) \equiv 0$  we have the simple result
\begin{equation}
\widetilde{G}_{QQ}(k)= \frac{\rho Z k^2}{k^2 + \kappa_{D}^{2}} \; (\rm{for}\; \rm{all}\;  k) \; ,
\end{equation}
and, in this case, the second SL rule is satisfied even at the CP.

Our last comment will be for the charge-density correlation function, which along the critical isochore above $T_{c}$  behaves as
\begin{equation}
\widetilde{G}_{NQ}(k)=  - \frac{\widetilde{\Phi}_{NQ}(0) }{4 \pi q^{2}} \frac{k^{2}}{a
(k ^{2} + \xi^{-2}) } +
\mathcal{O}(k^{4}) \; .
\end{equation}
at small $k$.
Therefore, for the asymmetric model $\widetilde{G}_{NQ}(k)\sim k^{2}$ away from the CP when $k\rightarrow 0$ and  $\widetilde{G}_{NQ}(k=0)=-\widetilde{\Phi}_{NQ}(0)/(4 \pi q^{2}a)\neq 0$ at the CP. Note that for the symmetric model
$\widetilde{G}_{NQ}\equiv 0$.

In summary, we have studied density-density, charge-charge and charge-density correlation functions for the charge- and size-asymmetric model far and at the gas-liquid CP within the framework of the RPA and have shown that in a general case the second-moment SL rule is satisfied away from CP and is not satisfied at the CP. In the particular case of the symmetrical model the both SL rules are satisfied at the CP. It is worth noting that a more comprehensive study  of the criticality in ionic fluids requires the  fluctuations to be taken into consideration at the level higher than the RPA. This issue as well as a  detailed analysis of an effect of the charge- and size-asymmetry on the phase behaviour  will be presented elsewhere.


\end{document}